\documentclass[sigplan,screen]{acmart}

\copyrightyear{2025}
\acmYear{2025}
\setcopyright{rightsretained}
\acmConference[ASPLOS '25]{Proceedings of the 30th ACM International Conference on Architectural Support for Programming Languages and Operating Systems, Volume 1}{March 30--April 3, 2025}{Rotterdam, Netherlands}
\acmBooktitle{Proceedings of the 30th ACM International Conference on Architectural Support for Programming Languages and Operating Systems, Volume 1 (ASPLOS '25), March 30--April 3, 2025, Rotterdam, Netherlands}
\acmDOI{10.1145/3669940.3707282}
\acmISBN{979-8-4007-0698-1/25/03}

\makeatletter
\gdef\@copyrightpermission{
  \begin{minipage}{0.3\columnwidth}
   \href{https://creativecommons.org/licenses/by/4.0/}{\includegraphics[width=0.90\textwidth]{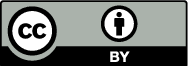}}
  \end{minipage}\hfill
  \begin{minipage}{0.7\columnwidth}
   \href{https://creativecommons.org/licenses/by/4.0/}{This work is licensed under a Creative Commons Attribution International 4.0 License.}
  \end{minipage}
  \vspace{5pt}
}
\makeatother

\usepackage[]{hyperref}
\usepackage{tabularx}
\usepackage{amsmath}
\usepackage{array}
\usepackage{makecell}
\usepackage{graphicx}
\usepackage{subcaption}
\usepackage{multirow}

\begin{document}

\title{Tally: Non-Intrusive Performance Isolation for Concurrent Deep Learning Workloads}

\makeatletter
\renewcommand{\@fnsymbol}[1]{\ifcase#1 \or †\else \@ctrerr\fi}
\makeatother

\author{Wei Zhao}
\authornote{Part of the work done as a student at the University of Toronto.}
\affiliation{
  \institution{Stanford University}
  \city{Stanford}
  \country{USA}
}
\affiliation{
  \institution{CentML}
  \city{Toronto}
  \country{Canada}
}
\email{wzhao@cs.stanford.edu}

\author{Anand Jayarajan}
\affiliation{
  \institution{University of Toronto}
  \city{Toronto}
  \country{Canada}
}
\affiliation{
  \institution{Vector Institute}
  \city{Toronto}
  \country{Canada}
}
\affiliation{
  \institution{CentML}
  \city{Toronto}
  \country{Canada}
}
\email{anandj@cs.toronto.edu}

\author{Gennady Pekhimenko}
\affiliation{
  \institution{University of Toronto}
  \city{Toronto}
  \country{Canada}
}
\affiliation{
  \institution{Vector Institute}
  \city{Toronto}
  \country{Canada}
}
\affiliation{
  \institution{CentML}
  \city{Toronto}
  \country{Canada}
}
\email{pekhimenko@cs.toronto.edu}

\renewcommand{\shortauthors}{Wei Zhao, Anand Jayarajan, and Gennady Pekhimenko}
%% No italics, no superscripts
%% Use footnote or author note to identify equal contribution and/or contact author info

\begin{CCSXML}
<ccs2012>
   <concept>
       <concept_id>10010147.10011777.10011778</concept_id>
       <concept_desc>Computing methodologies~Concurrent algorithms</concept_desc>
       <concept_significance>500</concept_significance>
       </concept>
   <concept>
       <concept_id>10010147.10010178</concept_id>
       <concept_desc>Computing methodologies~Artificial intelligence</concept_desc>
       <concept_significance>100</concept_significance>
       </concept>
   <concept>
       <concept_id>10010520.10010521.10010528</concept_id>
       <concept_desc>Computer systems organization~Parallel architectures</concept_desc>
       <concept_significance>300</concept_significance>
       </concept>
 </ccs2012>
\end{CCSXML}

\ccsdesc[500]{Computing methodologies~Concurrent algorithms}
\ccsdesc[100]{Computing methodologies~Artificial intelligence}
\ccsdesc[300]{Computer systems organization~Parallel architectures}

\keywords{GPU sharing, performance isolation, cloud infrastructure, deep learning, systems for machine learning}

\begin{abstract}
GPU underutilization is a significant concern in many production deep learning clusters, leading to prolonged job queues and increased operational expenses. A promising solution to this inefficiency is GPU sharing, which improves resource utilization by allowing multiple workloads to execute concurrently on a single GPU. However, deploying GPU sharing in production settings faces critical obstacles due to the limitations of existing mechanisms, including high integration costs, inadequate performance isolation, and limited application compatibility. To address these issues, we introduce \emph{Tally}, a non-intrusive GPU sharing mechanism that provides robust performance isolation and comprehensive workload compatibility. The key to Tally's robust performance isolation capability lies in its fine-grained thread-block-level GPU kernel scheduling strategy, which allows the system to effectively mitigate interference caused by workload co-execution. We evaluate Tally on a diverse range of workloads and show that it incurs an average overhead of only $7.2\%$ on the $99^{th}$-percentile latency of high-priority inference tasks when executed concurrently with best-effort training workloads, compared to $188.9\%$ overhead exhibited by the state-of-the-art GPU sharing systems like TGS, while achieving over $80\%$ of TGS's system throughput.
\end{abstract}

\maketitle % should come after the abstract
\pagestyle{fancy} % should come right after \maketitle

\section{INTRODUCTION}
Deep learning (DL) has demonstrated astonishing performance across a wide range of applications, including image recognition~\cite{krizhevsky2012, kaiming2016, szegedy2015}, machine translation~\cite{devlin-etal-2019-bert, bahdanau2015neural, sutskever2014, lewis-etal-2020-bart}, recommendation systems~\cite{Zhang2020, alashkar2017examples}, and autonomous driving ~\cite{Cordts2016, yurtsever2020survey}. Recent advances in generative AI~\cite{rombach2022high, brown2020language}, exemplified by the development of large language models (LLMs) such as ChatGPT~\cite{openai2024chatgpt}, have further elevated the influence of DL to unprecedented levels. To facilitate the research and deployment of DL-based applications, many organizations~\cite{shukla2022singularity, Weng2022mlaas, meta2024} are investing in building computing infrastructures with large fleets of GPU devices. For example, Meta has built two data center-scale clusters for AI, each equipped with hundreds of thousands of NVIDIA H100 GPUs with an estimated cost as high as half a billion dollars~\cite{meta2024, Eadline_2023}. As the demand for computational resources continues to escalate, efficient utilization of GPUs becomes paramount to mitigate costs and ensure scalability for AI applications.

Nevertheless, studies have noted significant GPU underutilization in large-scale DL clusters. For example, a recent study by Alibaba revealed that the median GPU utilization in their production DL cluster was only $4.2\%$~\cite{Weng2022mlaas}. This inefficiency arises from the need for production clusters to accommodate a diverse range of workloads, many of which are unable to fully utilize the capacity of GPUs. Training workloads, for instance, may underutilize GPUs due to bottlenecks in CPU execution such as stalls caused by data fetching and preprocessing~\cite{Audibert2023}. On the other hand, the utilization of online inference tasks is intrinsically tied to real-time request rates, which are subject to substantial fluctuations causing unpredictable GPU idleness~\cite{li2023-alpaserver}. For example, Microsoft's production trace~\cite{maf2} has revealed significant burstiness in user requests, with frequent spikes in demand up to $50\times$ the average. This makes usage prediction and resource allocation challenging for inference workloads, resulting in resource over-provisioning to ensure high quality of service~\cite{MArk2019, li2023-alpaserver}, which exacerbates the underutilization problem.

To enhance the efficiency of DL clusters, recent studies~\cite{xiao2020, salus2020, lucid, narayanan2020heterogeneity} suggest GPU sharing as a promising solution. This approach aims to enhance hardware utilization by allowing multiple low-utilization tasks to share a single GPU's resources. The same Alibaba study conducted a simulation on their production traces and demonstrated that an effective GPU sharing mechanism could reduce the amount of GPU resources required by the cluster by $50\%$ on average, with reductions reaching up to $73\%$ during peak demand periods. These findings underscore the potential of GPU sharing to mitigate GPU underutilization and significantly improve the efficiency of large-scale DL clusters.

Despite these promising results, GPU sharing has not yet achieved widespread adoption in production. One major obstacle is the lack of robust performance isolation mechanisms in most current GPU sharing systems. Many workloads in DL clusters are governed by service-level agreements (SLAs) with strict performance requirements, often measured in milliseconds~\cite{clipper2017}. These workloads are commonly referred to as \emph{high-priority} or \emph{latency-critical} tasks. In contrast, workloads that can tolerate extended processing times are known as \emph{low-priority} or \emph{best-effort} tasks. While it seems logical to pack best-efforts tasks along with high-priority tasks to utilize potential GPU idle time, this approach can cause significant interference to the performance of high-priority workloads, violating their SLAs. Our experiments on state-of-the-art GPU sharing mechanisms reveal that such interference can cause up to a $20\times$ slowdown in tail latency (see Section~\ref{sec:eval}), making them impractical for production.

Another hindrance to the broader adoption of GPU sharing lies in the intrusive nature of many existing solutions ~\cite{salus2020, xiao2020, case2022, effisha2017, slate2019, tacker2022}. These systems often necessitate extensive modifications to user code or the underlying DL frameworks (e.g., PyTorch ~\cite{paszke2019pytorch}, TensorFlow~\cite{tensorflow2016}). Such requirements not only greatly limit users' flexibility in developing workloads but also impose substantial burdens on users and cluster providers for workload adaptation and framework upgrades, leading to considerable integration and maintenance costs~\cite{lucid}. Moreover, many of these solutions require the workloads to adhere to specific characteristics, such as ensuring GPU kernels used by the workload are always idempotent~\cite{reef2022}. These constraints significantly limit their applicability in production clusters, which typically host a diverse range of workloads with heterogeneous characteristics.

In light of these observations, we argue that for practical deployment in production DL clusters, a GPU sharing system must meet the following three criteria: (1) it must be \emph{non-intrusive} to seamlessly integrate into existing clusters and DL workloads, (2) provide \emph{performance isolation} guarantees to ensure that the SLAs of high-priority tasks are effectively maintained during shared execution, and (3) be \emph{generalizable} across a diverse range of DL applications. Building on these objectives, we propose \emph{Tally}, a novel non-intrusive GPU sharing system for large-scale production DL clusters that ensures robust performance isolation for high-priority tasks when co-located with best-effort workloads, while maximizing hardware efficiency and overall system throughput. We design Tally as a transparent virtualization layer between the application level and the GPU. It intercepts GPU kernel launch API calls and employs a task-agnostic scheduling strategy that prioritizes the execution of high-priority kernels while opportunistically scheduling best-effort kernels during GPU idle periods.

To efficiently co-locate high-priority and best-effort tasks on a GPU with performance isolation guarantees, it is essential to schedule best-effort kernels in a controlled manner when the GPU is idle and promptly switch to high-priority kernels when needed. However, modern GPUs lack kernel scheduling primitives to terminate a running kernel and only allow context switching between kernels at the granularity of entire kernels~\cite{reef2022, GILMAN2021102234}. This limitation introduces significant overhead on high-priority tasks, especially since many best-effort DL workloads involve long-running kernels. To overcome this challenge and enable finer-grained kernel scheduling, Tally uses two block-level kernel scheduling primitives: \emph{slicing} and \emph{preemption}. Slicing divides large kernels into smaller segments, while preemption allows for the interruption of active kernels. Our key insight is that these two scheduling primitives can be implemented in a non-intrusive and task-agnostic manner through a series of GPU kernel transformation passes, leveraging the GPU programming model which organizes kernels as collections of independent thread blocks.

While slicing and preemption are effective mechanisms to manage co-execution interference, each has its trade-offs. Kernel slicing is a lightweight transformation but can incur higher overhead due to multiple kernel launches. In contrast, kernel preemption requires only a single kernel launch but involves a more complex transformation with potential synchronization overhead. Due to these differences, the optimal choice of scheduling primitives and configuration parameters (e.g., the degree of slicing) varies across different kernels. Towards this end, Tally uses a profile-guided resource provisioning strategy in its scheduling algorithm that transparently profiles the performance of best-effort kernels on-the-fly across various scheduling configurations to select the best option that meets the latency requirements of the high-priority task. This approach ensures that interference from best-effort tasks remains within acceptable limits while striving for optimal system throughput.

To evaluate Tally’s ability to provide effective performance isolation, we prepare a comprehensive benchmark suite consisting of six training and six inference workloads that cover a wide range of popular DL models and frameworks (details shown in table~\ref{benchmark-table}). We evaluate the co-located performance of latency-critical inference tasks and best-effort training tasks by simulating the production traffic patterns for the inference tasks using the publicly available Microsoft Azure Function Trace 2021 (MAF2) dataset~\cite{maf2}. Across all the combinations, Tally incurs a mere $7.2\%$ overhead on average in the $99^{th}$-percentile latency of inference tasks, compared to $252.3\%$, $345.0\%$, $195.5\%$, and $188.9\%$ exhibited by state-of-the-art non-intrusive GPU sharing solutions, Time-Slicing~\cite{GILMAN2021102234}, MPS, MPS-Priority~\cite{nvidia-mps}, and TGS~\cite{tgs2023} respectively. At the same time, Tally is able to achieve $80\%$ of the system throughput attained by TGS, demonstrating the effectiveness of Tally in maximizing the GPU efficiency while providing strong performance isolation guarantees.

In summary, we make the following contributions.
\begin{itemize}
\item We propose Tally, a non-intrusive GPU sharing solution with robust performance isolation guarantees.
\item We introduce a novel approach to achieve fine-grained GPU kernel scheduling in a non-intrusive and task-agnostic manner using kernel slicing and preemption primitives. We further propose a priority-aware kernel scheduler that dynamically selects the appropriate scheduling primitives to optimize GPU throughput while upholding performance isolation guarantees.
\item We evaluate Tally on a diverse range of training and inference workloads and show that Tally can achieve system throughput on-par with state-of-the-art GPU sharing solutions while ensuring strong performance isolation.
\end{itemize}
\section{BACKGROUND}\label{sec:bac}
Modern large-scale DL clusters are commonly shared among numerous users within an organization and are designed to support a diverse range of workloads~\cite{xiao2020, lucid}. These include training, fine-tuning, and inference tasks for a variety of model architectures, such as convolutional neural networks (CNNs)~\cite{Qi_2017_CVPR, kaiming2016}, transformer-based large language models (LLMs)~\cite{touvron2023llama, gpt-neo, devlin-etal-2019-bert}, graph neural networks (GNNs)~\cite{graph-transformer2019, graph-attention2018}, and reinforcement learning (RL) models~\cite{schulman2017proximal, Silver2016}. These workloads are implemented using various DL frameworks, including widely adopted open-source platforms like PyTorch~\cite{paszke2019pytorch}, TensorFlow~\cite{tensorflow2016}, and JAX~\cite{jax2018github}, as well as proprietary software stacks~\cite{Weng2022mlaas}. Given the diversity and compute-intensive nature of DL workloads, general-purpose GPUs have become the predominant accelerator in DL clusters for their expressive programming model and massive parallel processing capabilities.

\begin{figure}[t]
  \centering
  \includegraphics[width=0.9\linewidth]{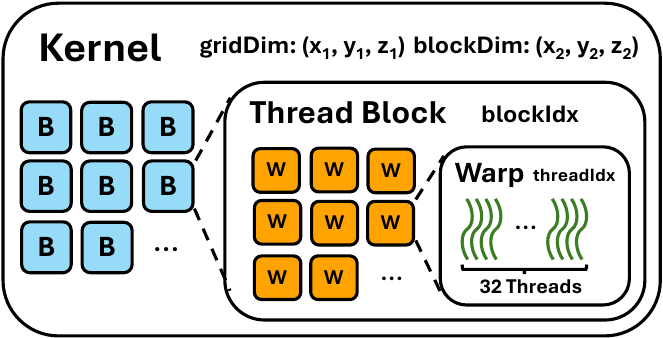}
  \caption{Overview of the GPU programming model.}
  \label{fig:gpu-model}
\end{figure}

\noindent \textbf{GPU architecture and programming model.} Despite variations across vendors (e.g, NVIDIA, AMD), the core architecture and programming model of modern GPUs in DL clusters remain largely similar. GPUs are composed of hundreds of parallel processing units called Streaming Multiprocessors (SMs). The computational tasks executed on GPUs, known as kernels, are programmed using specialized languages like CUDA~\cite{cuda2024}. As illustrated in Figure~\ref{fig:gpu-model}, GPU kernels follow a hierarchical thread group structure consisting of threads, warps, thread blocks, and grids. Threads within a kernel identify themselves using built-in variables such as \textit{threadIdx} and \textit{blockIdx} to determine their task assignments. Threads within a warp execute instructions in a SIMT (Single Instruction, Multiple Threads) fashion, where multiple threads execute the same instruction on different pieces of data simultaneously. All warps within a thread-block are scheduled onto a single SM, enabling them to share data using fast on-chip shared memory. Additionally, threads within a block can synchronize their execution using synchronization primitives to coordinate memory accesses. Finally, thread blocks within a grid can be scheduled and executed independently in any order, either sequentially or in parallel~\cite{cuda2024}. This independent scheduling capability allows thread blocks to scale effectively across multiple SMs.

\noindent \textbf{GPU sharing systems.} The diverse nature of DL workloads, coupled with their inherent inefficiencies, has led to significant concerns about GPU underutilization in many production DL clusters~\cite{hu2021characterization, xiao2020}. To address this challenge, researchers have proposed numerous cluster-level optimizations~\cite{xiao2018, xiao2020, wang2021horizontally, salus2020}, among which GPU sharing has emerged as a promising approach to improve utilization by consolidating multiple low-utilization jobs onto the same GPU. From an implementation perspective, current GPU sharing solutions can be broadly categorized into two levels: application-level and runtime-level approaches. Application-level solutions typically involve the introduction of new programming interfaces designed for concurrent DL execution. For example, Salus~\cite{salus2020} exposes a set of scheduling primitives for defining the iteration boundaries of a training job, enabling the scheduler to manage different training workloads at the iteration-level granularity. Similarly, HFTA~\cite{wang2021horizontally} provides APIs that enable developers to group multiple hyper-parameter tuning jobs into a single application, facilitating efficient execution on a single GPU.

In contrast to application-level solutions that require modifications to user-level applications, runtime-level GPU sharing solutions operate transparently by integrating directly into either DL frameworks~\cite{xiao2020} (e.g., PyTorch~\cite{paszke2019pytorch}, TensorFlow~\cite{tensorflow2016}), container orchestration systems~\cite{tgs2023} (e.g., Kubernetes~\cite{kubernetes}), or the GPU driver~\cite{reef2022, nvidia-mps, nvidia-mig}. For example, NVIDIA GPUs provide two built-in driver-level concurrency mechanisms: Time-Slicing~\cite{GILMAN2021102234} and MPS~\cite{nvidia-mps}, which enable transparent sharing of GPU resources in temporal and spatial manners respectively. Additionally, NVIDIA's Multi-Instance GPU (MIG)~\cite{nvidia-mig} introduces hardware-level resource partitioning primitives that divide GPU resources, such as SMs, cache, and memory bandwidth, into multiple virtual GPU instances, each operating within an isolated environment. Besides vendor-provided solutions, TGS~\cite{tgs2023} integrates with Kubernetes to schedule workloads at the GPU kernel level, while REEF~\cite{reef2022} extends the GPU driver to offer more fine-grained thread-level scheduling capabilities. The transparent design of these solutions facilitates broader adoption across applications. However, designing these solutions is often challenging due to the lack of application-level context.
\section{GPU SHARING IN THE WILD}
\label{sec:motivation}

Despite the wide range of solutions, current GPU sharing systems are inadequate for practical deployment in DL clusters due to one or more of the following key limitations.

\noindent\textbf{High integration and maintenance cost.} Many existing GPU sharing systems are fundamentally intrusive in nature, necessitating extensive modifications to user code to be used in practice. This is particularly evident in application-level systems, where users must adapt their workloads to the specific programming interfaces proposed by these solutions. Such intrusive approaches not only burden the users, but also limit their flexibility in application development~\cite{lucid}. Similarly, runtime-level GPU sharing solutions that are designed as extensions of DL frameworks impose a significant burden on cluster managers to implement and maintain them across diverse frameworks used by the users. Consequently, deploying these solutions in practice often involves substantial integration and maintenance costs, especially given the rapidly evolving landscape of modern DL ecosystem.

On the other hand, while runtime-level GPU sharing solutions that integrate seamlessly with container orchestration systems or the GPU driver may appear non-intrusive, they often impose implicit requirements on user code to provide additional application contexts needed for making effective scheduling decisions. These include the ability to temporarily launch applications in a specialized profiling environment prior to full execution~\cite{orion2024}, or mandating applications to implement checkpoint and restart mechanisms to support live migration~\cite{gavel2020, xiao2018}. However, enforcing such programming practices in large-scale, multi-tenant clusters is often impractical due to the operational complexity it introduces~\cite{Weng2022mlaas}.

\begin{table}[t]
\centering
\caption{Comparison of BERT inference latency against turnaround latency of different scheduling granularity for Whisper training on NVIDIA A100 GPU.}
\begin{tabular}{|c|cccc|}
\hline
\multirow{2}{*}{\begin{tabular}[c]{@{}c@{}}Inference time\\ (BERT)\end{tabular}} & \multicolumn{4}{c|}{Turnaround latency (Whisper)}                                                                   \\ \cline{2-5} 
& \multicolumn{1}{c|}{Iteration} & \multicolumn{1}{c|}{Kernel}      & \multicolumn{1}{c|}{Block}        & Thread      \\ \hline
$3.93 ms$                                                                          & \multicolumn{1}{c|}{$\sim3s$} & \multicolumn{1}{c|}{$\sim10ms$} & \multicolumn{1}{c|}{$\sim304\mu s$} & $\sim38\mu s$ \\ \hline
\end{tabular}
\label{tab:scheduling-granularity}
\end{table}

\noindent\textbf{Lack of performance isolation guarantees.} In production DL clusters, many inference tasks are governed by service-level agreements (SLAs), which impose strict stringent requirements on their response times, typically ranging from a few milliseconds to seconds~\cite{clipper2017}. However, current GPU sharing solutions often fall short in providing adequate performance isolation, leading to severe performance degradation for latency-critical inference tasks when co-located with other workloads~\cite{GILMAN2021102234}. For instance, NVIDIA MPS~\cite{nvidia-mps} is designed to eagerly schedule as many GPU kernels from co-located tasks as possible. While this design maximizes GPU utilization, it can cause significant queuing delays for executing kernels of the high-priority inference tasks. Our experiments (see Section~\ref{sec:eval}) reveal that the tail latency of a co-located inference task under MPS can increase up to $20\times$ compared to their standalone performance.

To effectively share GPU resources while ensuring performance isolation, a GPU sharing system must prioritize one task over another and rapidly switch between them according to their dynamic resource demands. The speed of context-switching between tasks, which we denote as the \emph{turnaround latency}, is dictated by the scheduling granularity employed by the GPU sharing system. As described in Section~\ref{sec:bac}, the scheduling granularity used by prior works primarily fall into four categories, from coarsest to finest: iteration-level (e.g. Salus~\cite{salus2020}), kernel-level (e.g., TGS~\cite{tgs2023}), block-level (e.g. EffiSha~\cite{effisha2017}), and thread-level (e.g. REEF~\cite{reef2022}). To illustrate the effect of scheduling granularity on the tail latency, we present a representative study in Table~\ref{tab:scheduling-granularity} where a latency-critical BERT~\cite{devlin-etal-2019-bert} inference task is co-executed with a Whisper~\cite{radford2023robust} model training workload on an NVIDIA A100 GPU~\cite{nvidia_a100}. In this experiment, we empirically estimate the turnaround latency for different scheduling granularities on the Whisper model and compare it to the response time of BERT inference. Since a BERT inference request takes only $3.93$ ms, achieving strong performance isolation requires turnaround latencies to be in the sub-millisecond range. However, both iteration-level and kernel-level scheduling can only achieve turnaround latencies of $3$ s and $10$ ms, respectively. As a result, scheduling the Whisper training task at these granularities can introduce $2.5-760\times$ higher overhead on BERT's inference latency. In contrast, block-level and thread-level scheduling achieve microsecond-scale turnaround latencies of $304$ µs and $38$ µs, respectively, resulting in only mere $1-7\%$ overhead on the latency.

From these results, we conclude that to achieve practical performance isolation guarantees, GPU sharing solutions must schedule resources at either the block-level or thread-level granularity. Unfortunately, existing solutions either do not meet these criteria (e.g., Salus~\cite{salus2020}, AntMan~\cite{xiao2020}, TGS~\cite{tgs2023}) or prove ineffective in diverse production DL clusters due to their intrusive nature (e.g., EffiSha~\cite{effisha2017}, Orion~\cite{orion2024}) or lack of generalizability (e.g. REEF~\cite{reef2022}). This brings us to the third limitation.

\noindent\textbf{Reliance on narrowly applicable workload characteristics.} Many state-of-the-art GPU sharing systems rely on specific workload characteristics in their scheduling strategies. For instance, AntMan and TGS provide runtime-level GPU sharing support but are limited to training workloads with predictable GPU kernel execution patterns. REEF, on the other hand, offers strong performance isolation for inference workloads; however, it requires all GPU kernels to be idempotent to support thread-level kernel scheduling strategies, meaning the kernels must always produce the same output for a given input, regardless of how many times it is executed. While these systems demonstrate benefits for certain classes of applications, their underlying assumptions are often not applicable to large-scale multi-tenant DL clusters that must support a diverse range of workloads with varying characteristics. More critically, prior studies~\cite{Weng2022mlaas} have revealed that DL clusters are often agnostic to the specific semantics of running workloads and extracting application-level context is largely infeasible. As a result, GPU sharing solutions that depend on specific task characteristics are not viable in many DL clusters.

In summary, we argue that a practical GPU sharing solution for large-scale DL clusters must address all three of the aforementioned limitations.

\section{TALLY: SYSTEM DESIGN}
\label{sec:tally-system}
To facilitate the practical deployment of GPU sharing in production DL clusters, we propose \emph{Tally}, a novel non-intrusive GPU sharing system that provides robust performance isolation for high-priority tasks during shared execution with best-effort tasks across a diverse range of DL workloads. Tally is designed as a virtualization layer that operates transparently between applications and the GPU, intercepting device API calls and employing a task-agnostic scheduling algorithm to \emph{prioritize} execution from \emph{high-priority tasks} while \emph{opportunistically scheduling} kernels from \emph{best-effort tasks} during GPU idle cycles.

Tally ensures performance isolation for high-priority tasks by scheduling kernels from best-effort tasks at the block-level granularity using two scheduling primitives: slicing and preemption. Slicing divides a large kernel into numerous smaller segments, enabling more controlled scheduling, whereas preemption allows for the interruption of active kernels to expedite resource reallocation. Importantly, we make two key observations that allow applying these two scheduling primitives in a task-agnostic and non-intrusive manner. Firstly, unlike thread-level scheduling, thread blocks within a GPU are guaranteed to operate independently, making block-level scheduling universally applicable across GPU kernels. Moreover, while thread-level scheduling can provide stronger performance isolation, we find that block-level scheduling is sufficient for most practical scenarios. Secondly, the slicing and preemption scheduling primitives can be applied on GPU kernels through a series of transformation passes on the kernel device code (PTX code in the case of NVIDIA GPUs) obtained through the interception of device API calls. Leveraging these insights, we integrate the slicing and preemption primitives to Tally's scheduling algorithm to enable microsecond-scale turnaround latency, allowing Tally to promptly switch execution between high-priority and best-effort tasks and maintain the millisecond-scale latency requirements in many latency-critical tasks.

\begin{figure}[t]
  \centering
  \includegraphics[width=0.90\linewidth]{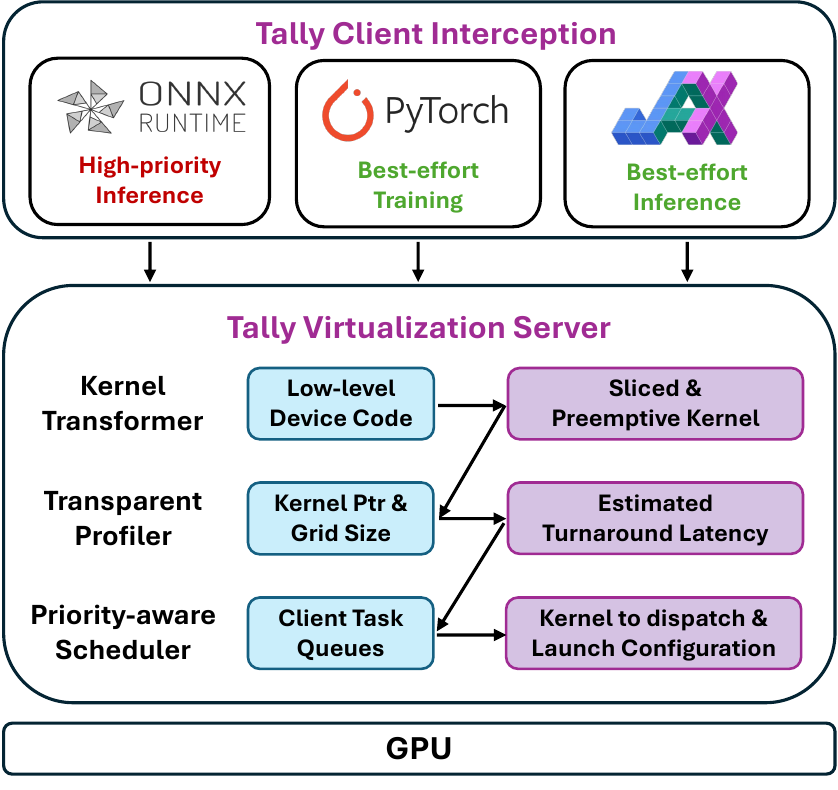}
  \caption{Tally architecture.}
  \label{fig:tally-arch}
  \vspace{-10pt}
\end{figure}

\begin{figure*}[t]
    \centering
    \begin{subfigure}[b]{0.45\textwidth}
        \centering
        \includegraphics[width=\textwidth]{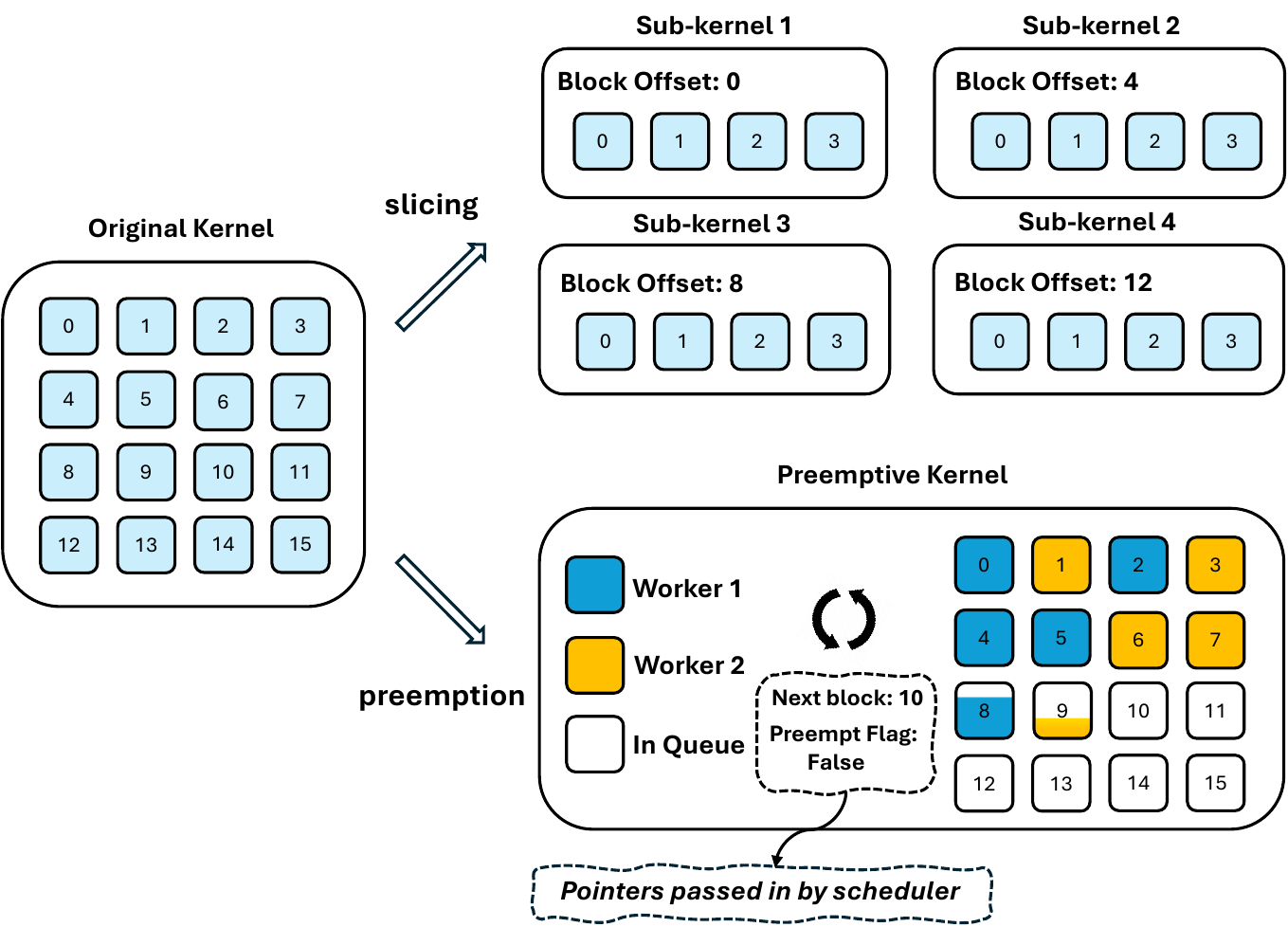}
        \label{fig:transformation}
    \end{subfigure}
    \hfill
    \begin{subfigure}[b]{0.51\textwidth}
        \centering
        \includegraphics[width=\textwidth]{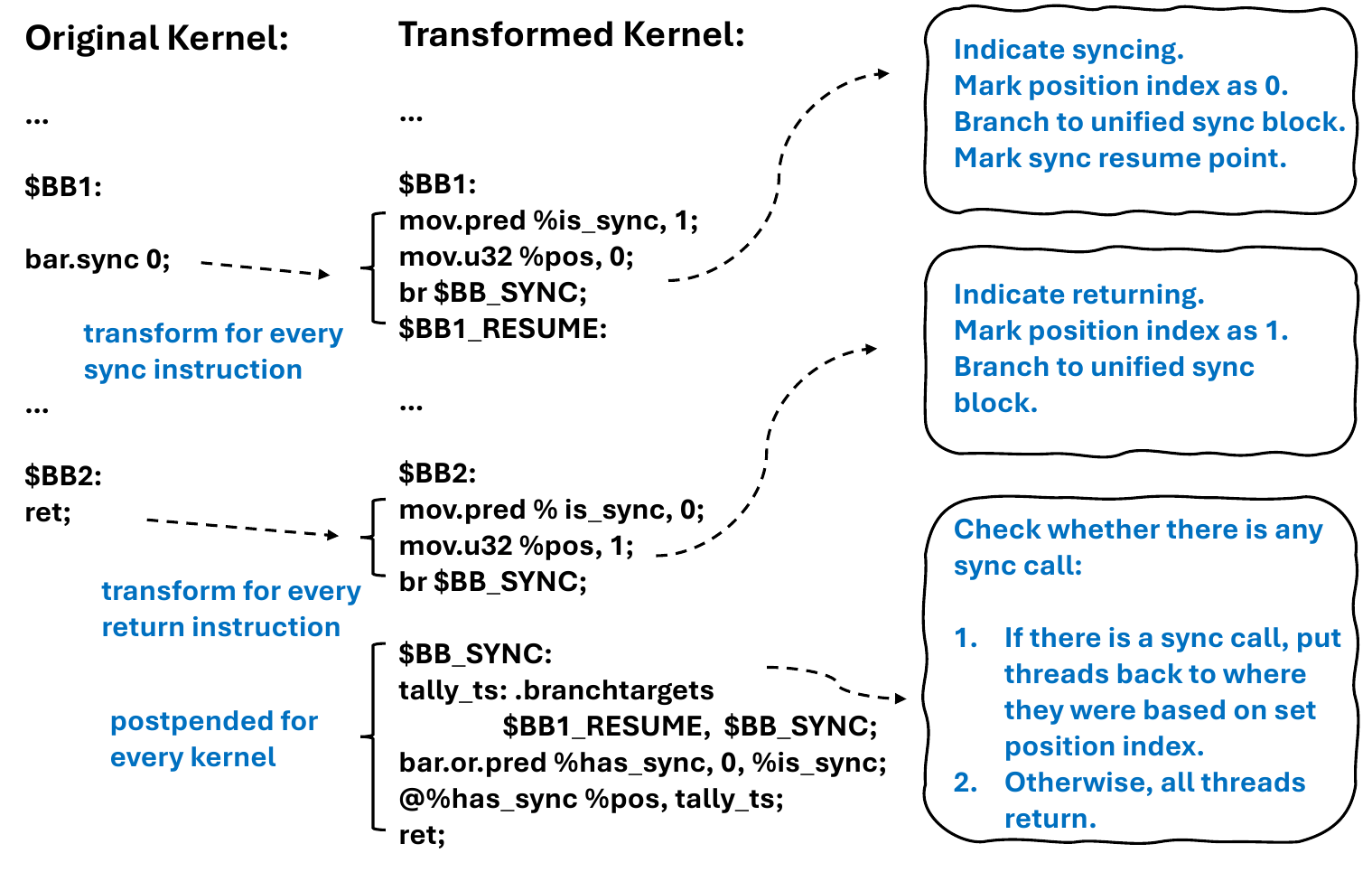}
        \label{fig:unified-transformation}
    \end{subfigure}
    \caption{Kernel transformations: (a) slicing and preemption transformations (b) unified synchronization transformation.}
    \label{fig:transformation-combined}
\end{figure*}

Figure~\ref{fig:tally-arch} shows the high-level design of Tally, which follows a client-server architecture. On the application side, Tally's client library intercepts device code and API calls initiated by client processes and forwards them to the Tally server, which is responsible for managing the actual device execution and resource scheduling. The objective of the Tally server is to facilitate priority-enforced concurrent execution of one high-priority task alongside multiple best-effort tasks. To achieve this, the Tally server incorporates three key components: a (1) \textit{kernel transformer}, which leverages intercepted device code to convert kernels into slicing and preemptive-style kernels, a (2) \textit{transparent profiler}, which profiles the performance of the transformed kernels under various launch configurations at runtime and estimates their turnaround latencies, and finally a (3) \textit{priority-aware scheduler}, which integrates capabilities of the aforementioned modules and strategically schedules GPU execution in a priority-enforced manner. In the following sections, we provide details regarding each of these components.

\subsection{Kernel Transformations}
\label{sec:transformations}
We first illustrate how Tally supports the slicing and preemption primitives through a series of transformation passes on the kernel device code.

\noindent \textbf{Slicing transformation.} Slicing is a technique that facilitates fine-grained control over kernel execution by partitioning the thread blocks of a kernel into multiple sub-kernels. Figure~\ref{fig:transformation-combined}(a) illustrates this concept, where a kernel originally composed of $16$ thread blocks is divided into $4$ sub-kernels, each containing $4$ blocks. As thread blocks within a kernel are guaranteed to be independent, theoretically, this partitioning should preserve the original kernel's functional semantics. However, naively launching the original kernel $4$ times with $4$ thread blocks each will yield incorrect results, as GPU threads rely on built-in variables such as \textit{blockIdx} to determine their task assignment. Specifically, the \textit{blockIdx} variable informs a thread of the index of the block it resides in within the kernel execution, allowing it to properly compute its assigned task index. In the original kernel, threads are exposed to \textit{blockIdx} values ranging from $0$ to $15$ (as $16$ blocks in total are launched). Yet, when each sub-kernel is launched, the threads are exposed to \textit{blockIdx} values ranging from $0$ to $3$, causing all sub-kernels to execute the first $4$ blocks of the kernel, which will lead to erroneous results. To address this issue, we modify a kernel's device code to take in an additional offset variable in its parameters and substitute occurrences of \textit{blockIdx} with the sum of the offset and the original \textit{blockIdx}. This approach enables threads within a sub-kernel to identify their corresponding subset of blocks. For instance, sub-kernel $2$ can reconstruct its corresponding block indices from $4$ to $7$ by adding a block offset of $4$ to the new \textit{blockIdx}. With this mechanism, the collective work done by the sub-kernels can properly resemble the original kernel execution, while offering the scheduler flexible scheduling at the block-level granularity.

\noindent \textbf{Preemption transformation.} Tally introduces a transformation that makes kernels preemptible while preserving their functional semantics. The preemption transformation in Tally is inspired by the Persistent Thread Block (PTB) programming paradigm~\cite{ptb}, which deviates from the traditional programming model where the number of thread blocks launched by a kernel is proportional to the amount of work it performs. Instead, the PTB model employs a flexible number of worker blocks to perform all the work. As an example, in Figure \ref{fig:transformation-combined}(a), rather than launching $16$ thread blocks of the original kernel, the PTB model launches $2$ worker blocks to process the $16$ blocks in an iterative and dynamic manner. These workers continuously fetch and increment a global counter which represents the next task index. Similar to how sub-kernels in slicing utilize block offsets to recompute their block indices from the original kernel execution, workers in PTB use the task index to reconstruct the corresponding block indices for the current iteration. For example, in the illustrated case where thread blocks are structured in a $1$-dimensional grid, a fetched task index of $8$ indicates that the worker should perform the execution of the thread block with \textit{blockIdx} $8$. In the case of higher-dimensional grids, block indices can be computed from the integer task index by referring to the original kernel's grid dimension.

A critical observation from the PTB model is that the progress of the kernel execution is effectively encapsulated in the global task index. Since the execution is now transformed into an iterative approach, kernels can potentially terminate early and later be resumed from the point at which they ended, using the task index. This can be achieved by introducing an additional boolean variable to the PTB kernels as a flag for signaling preemption. At the start of each iteration, workers check this flag to detect preemption signals, thereby enabling block-level preemption within kernels. In Tally, we automatically convert kernels into this preemptive style by wrapping the original kernel function in the device code with an outer loop using control flow logic. For instance, all return statements are transformed into branch instructions that direct the execution flow to the start of the iteration. This ensures that each worker block will continue execution until all tasks are processed or a preemption signal is received. Furthermore, a synchronization call is introduced at the end of each iteration to make sure that threads within a worker block process a single block at a time, preventing potential race conditions and maintaining correctness of computation.

\noindent \textbf{Unified synchronization transformation.} In the preemption transformation, return statements are substituted with branching statements that redirect threads to synchronize at the start of the iteration. However, this modification introduces a risk of threads within the same block calling synchronization at different places within the kernel. For instance, while some threads may arrive at one of the kernel’s original synchronization points, others may be redirected to the newly introduced synchronization point due to early returning. Such divergence in synchronization is an undefined behavior in the GPU programming model and, in practice, results in infinite kernel stalls. This issue makes the preemption transformation unsafe to be applied on all kernels.

To address this problem, Tally introduces a prepositional kernel transformation pass which we denote as the \textit{unified synchronization transformation}. The algorithm for this transformation is depicted in Figure \ref{fig:transformation-combined}(b). As illustrated, this transformation modifies all synchronization and return instructions by redirecting them to a unified synchronization point that checks whether all threads have reached a return state. If this condition is met, all threads return simultaneously; otherwise, threads from synchronization points loop back to their previous execution positions, while those returning are held until all threads align. This modification guarantees that threads within a block return synchronously. Thus, when the kernel undergoes the preemption transformation, the risk of divergence in thread block synchronization due to early returns is eliminated, thus resolving the issue of infinite stalls. Through this prepositional transformation stage, the preemptive transformation can be applied automatically and safely to all kernels.

% \begin{figure}[t]
% \begin{minted}[
% linenos,
% fontsize=\footnotesize,
% xleftmargin=12pt,
% numbersep=5pt,
% ]{python}
% def launch_and_profile(kernel):
%     candidates = get_candidate_configs(kernel)
%     for cfg in candidates:
%         est_turnaround = lookup_measurement(kernel, cfg)
%         if est_turnaround is None:
%             est_turnaround = profile(kernel, cfg)
%         if has_launched(): 
%             return
%     set_launch_config(kernel, candidates, 
%                       bound=TUNEAROUND_LATENCY_BOUND)
% def scheduler():
%     while True:
%         for client in sort_by_priority(clients):
%             if client.is_high_priority():
%                 kernel = client.fetch_next_kernel()
%                 if kernel is not None:
%                     preempt_best_effort_kernel()
%                     kernel.launch(Config::DEFAULT)        
%                 if client.has_kernel_running():
%                     break
%             else:    
%                 kernel = client.get_curr_ex_kernel()
%                 if kernel is None:
%                     kernel = client.fetch_next_kernel()
%                     cfg = lookup_launch_config(kernel)
%                     if cfg is not None:
%                         kernel.launch(cfg)
%                     else:
%                         launch_and_profile(kernel)
%                 elif kernel.running():
%                     pass
%                 elif kernel.stopped():
%                     kernel.resume()
% \end{minted}
% % \vspace{-10pt}
% \caption{Tally's priority-aware scheduling algorithm.}
% \label{scheduling-algorithm}
% \end{figure}

\begin{figure}[t]
  \centering
  \includegraphics[width=\linewidth]{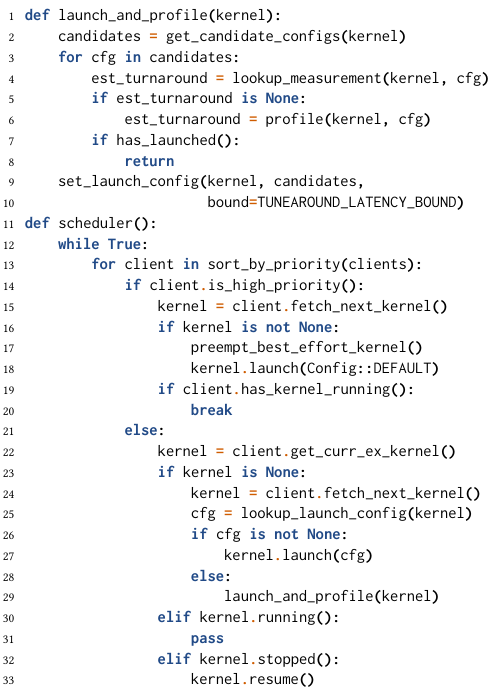}
  \caption{Tally's priority-aware scheduling algorithm.}
  \label{scheduling-algorithm}
\end{figure}

\subsection{Priority-aware Scheduling}

We now demonstrate how Tally's priority-aware scheduler leverages the aforementioned slicing and preemption primitives alongside the transparent profiler to enforce priority during the scheduling of concurrent workloads. The algorithm of the scheduler, as outlined in Figure~\ref{scheduling-algorithm}, follows an opportunistic strategy: it prioritizes the immediate dispatch of high-priority kernels upon their arrival (lines $14$ to $20$) and schedules low-priority kernels to execute only when the high-priority job becomes inactive (lines $21$ to $33$). Due to the reactive nature of this algorithm, incoming high-priority kernels may need to wait for the completion of currently executing low-priority kernels, potentially introducing delays to the execution of high-priority tasks. To mitigate such queuing delays, each best-effort kernel is either launched as multiple sub-kernels one at a time, or in its preemptive form, such that it can be preempted in the event of a high-priority kernel arriving during its execution.

The primary challenge of this strategy lies in determining the appropriate launch configuration for each best-effort kernel, which includes deciding whether to employ slicing or preemption and setting the corresponding parameters for each method. First, due to the inherent difference in their implementation, each of the two primitives exhibits distinct performance trade-offs. While slicing is a lightweight transformation, it can incur overhead due to multiple kernel launches. Conversely, the preemption kernel requires only a single launch, but the additional synchronization and control flow can introduce noticeable overhead to certain kernels. As a result, their relative performance across different kernels varies considerably, posing challenges for the scheduler in choosing between the two. Moreover, for each primitive, the scheduler need to determine the corresponding parameter, which includes the number of slices for slicing and the number of workers for preemption. These decisions are crucial as they significantly affect the performance behavior of the concurrent workloads. For instance, in the context of preemptive scheduling, allocating fewer workers can expedite the preemption process but can potentially slow down the execution of best-effort tasks, thereby diminishing the overall benefits of GPU sharing.

To strike a balance between ensuring robust performance isolation for high-priority tasks and maintaining acceptable throughput for low-priority tasks, Tally's priority-aware scheduler employs a profile-guided resource provisioning policy. This policy leverages runtime performance metrics to determine the optimal launch configuration for each best-effort kernel through a search process. As depicted in Figure~\ref{scheduling-algorithm}, lines $2$ to $10$, for each kernel, the scheduler generates a set of candidate configurations encompassing both slicing and preemption. For preemption, potential configurations are derived by considering all possible multiples of the number of SMs that fit with the thread constraints. Whereas slicing may set the degree of slicing based on having each slice operating on different percentages of the total blocks. These candidate configurations are then assessed by Tally's transparent profiler.

To evaluate the performance of candidate configurations, Tally employs the \emph{turnaround latency} metric, as introduced in Section~\ref{sec:motivation}. This metric represents the estimated time needed for a best-effort kernel to release occupied resources. Calculating the turnaround latency for a sliced kernel is relatively straightforward, which can be done by simply measuring the completion time of a single kernel slice. In contrast, determining the turnaround latency for preemptive kernels is more challenging due to their use of dynamic worker blocks. To address this challenge, we employ a heuristic approach that approximates the turnaround latency by dividing the total execution time of the kernel by the number of blocks allocated to each worker, as expressed by:\begin{equation} \textit{turnaround\_latency} = \frac{\textit{kernel\_latency} \cdot \textit{worker\_blocks}}{\textit{total\_blocks}}.\end{equation} This formula approximates the latency of waiting for each worker to complete the current block of execution. After profiling each configuration, the scheduler selects the one that achieves the optimal performance while complying with a predefined turnaround latency threshold, which serves as a configurable parameter to balance the performance trade-offs between high and low-priority workloads. Empirically, we find that a default value of $0.0316$ ms can effectively render robust performance isolation while producing competitive system throughput (details in Section~\ref{sec:turnaround_latency}).

Notably, Tally’s profiling process accounts for variations in kernel inputs by profiling each unique work configuration (block and grid dimensions) separately for every kernel. For measurement stability, profiling results are averaged across many runs (e.g., $10$). Since profiling operates on a per-kernel basis, once measurements for a kernel are collected, they can be reused throughout the execution, thus introducing negligible overhead to overall execution.

\subsection{Non-intrusive Virtualization Layer}
\label{virtualization-section}
Finally, we present details on the implementation of Tally's virtualization layer. On the application-side, Tally utilizes the \texttt{LD\_PRELOAD} mechanism in Linux to intercept device API calls, redirecting them to a surrogate function that mirrors the API's original signature. These intercepted API calls are sent to the server, which handles the actual device execution. The challenge in this process is enabling the server, which operates in a separate address space, to execute client kernel functions. The solution leverages a critical insight: GPU applications register device code with the GPU driver at the start of execution. By intercepting this process, the server gains access to the device code, including kernel binaries and PTX code, which we use at the server-side to apply kernel transformations and recompile to executable GPU kernels.

A notable concern with this interception and forwarding approach is the potential communication overhead, which could impact application performance. To mitigate this, we have implemented two critical optimizations. First, we utilize efficient shared memory channels for client-server communication, which significantly reduces the overhead by avoiding context switches during message passing. Second, we observe that the need to forward many frequent device API calls, such as \textit{cudaGetDevice} in NVIDIA, can be avoided by maintaining local states of the execution context on the client side. This approach effectively minimizes the frequency of API call forwarding between the client and server, further reducing the communication overhead. These strategies collectively enable Tally to operate transparently with near-native performance across all the concurrent DL tasks.
\section{EVALUATION}
\label{sec:eval}

\subsection{Methodology}

\noindent \textbf{Workloads.} To evaluate Tally, we prepare a benchmark suite consisting of six training and six inference workloads as listed in Table~\ref{benchmark-table}, covering a diverse range of widely used DL model architectures, including CNNs (e.g., ResNet50~\cite{kaiming2016}), transformer-based LLMs (e.g., Llama-2~\cite{touvron2023llama}), and diffusion models (e.g., Stable Diffusion~\cite{rombach2022high}). Training workloads are implemented using PyTorch~\cite{paszke2019pytorch}, a broadly adopted DL framework in the research community. For inference workloads, we use high-performance inference engines such as ONNX Runtime~\cite{onnxruntime}, TorchInductor~\cite{torchinductor2024}, and Hidet~\cite{hidet2023}. Some kernels in the workloads are sourced from proprietary libraries such as cuBLAS~\cite{cuBLAS}, which restrict device code interception. To enable block-level transformations, Tally automatically replaces these kernels at runtime with CUTLASS~\cite{cutlass} alternatives that offer similar performance. Inspired by prior DL serving benchmarks~\cite{li2023-alpaserver, reef2022, bhattacharjee2019barista}, we simulate inference workloads using the Microsoft Azure Function Trace 2021 (MAF2)~\cite{maf2}, a dataset of Azure serverless function invocations that has been extensively repurposed for DL serving research. We extract the trace of the most frequently invoked function and adapt it to match specific traffic loads based on the inference latency of the models. We define \emph{load} as the percentage of time an inference service is actively serving requests.

\noindent \textbf{Metrics.} We use the \emph{$99^{th}$ percentile latency} as the primary metric for assessing the performance of inference tasks. For both training and inference tasks, we measure their \textit{throughput}, defined as the number of samples processed per unit of time. To evaluate the overall performance of concurrent task execution, we calculate \textit{system throughput} by summing the normalized throughput of all concurrent workloads.

\noindent \textbf{Baselines.} We compare the performance of Tally against four state-of-the-art non-intrusive GPU sharing solutions: (i) Time-Slicing, (ii) MPS, (iii) MPS-Priority, and (iv) TGS. All these systems perform scheduling at the kernel-level granularity. Specifically, Time-Slicing and MPS are the proprietary concurrency mechanisms provided by NVIDIA GPUs. MPS-Priority refers to MPS with the client priority control feature enabled, which allows assigning priority values to workloads for prioritized resource allocation. TGS is a GPU sharing system specifically designed for DL workloads that leverages adaptive rate control mechanisms to achieve performance isolation across concurrent workloads.

\noindent \textbf{Experimental setup.} All experiments are conducted on an AWS EC2 p4d.24xlarge instance, a production-grade GPU server widely deployed in DL clusters. It is equipped with a $96$-core Intel Xeon Platinum 8275CL CPU, $1152$ GB of DRAM, and $8$ NVIDIA A100 SXM GPUs, each with $40$ GB of memory. The server runs on Ubuntu 20.04, and is configured with CUDA 12.2, PyTorch 2.2.0, ONNX Runtime GPU 1.17.0, and Hidet 0.3.0.

\renewcommand{\arraystretch}{1.2}
\begin{table}[t]
\footnotesize
\centering
\begin{tabularx}{\columnwidth}{>{\centering\arraybackslash}p{2.6cm} >{\centering\arraybackslash}p{1.9cm} >{\centering\arraybackslash}p{1.15cm} >{\centering\arraybackslash}p{1.6cm}}
% \Xhline{2\arrayrulewidth}
%\multicolumn{4}{c}{\textbf{Training Workloads}} \\
\Xhline{2\arrayrulewidth}
\textbf{Training Model} & \textbf{Dataset} & \textbf{\# Params} & \textbf{Throughput} \\
\hline
ResNet50~\cite{kaiming2016} & ImageNet~\cite{imagenet2009} & 25.6M & 1.0 it/s \\
PointNet~\cite{Qi_2017_CVPR} & ShapeNet~\cite{chang2015shapenet} & 3.5M & 40.0 it/s \\
BERT~\cite{devlin-etal-2019-bert} & SQuAD~\cite{rajpurkar-etal-2016-squad} & 110M & 1.8 it/s \\
GPT2-Large~\cite{radford2019language} & Wikitext2~\cite{merity2016pointer} & 774M & 3.3 it/s \\
PEGASUS~\cite{zhang2020pegasus} & XSum~\cite{xsum-emnlp} & 568M & 2.9 it/s \\
Whisper-v3~\cite{radford2023robust} & LibriSpeech~\cite{librispeech2015} & 1.5B & 0.3 it/s \\
\Xhline{2\arrayrulewidth}
\noalign{\vskip 3mm}
%\multicolumn{4}{c}{\textbf{Inference Workloads}} \\
\Xhline{2\arrayrulewidth}
\textbf{Inference Model} & \textbf{Engine} & \textbf{\# Params} & \textbf{Latency} \\
\hline
ResNet50~\cite{kaiming2016} & Hidet~\cite{hidet2023} & 25.6M & 1.37 ms \\ 
BERT~\cite{devlin-etal-2019-bert} & ONNX RT~\cite{onnxruntime} & 110M & 3.93 ms \\ 
YOLOv6m~\cite{li2022yolov6} & TorchInductor~\cite{torchinductor2024} & 34.9M & 17.5 ms\\ 
Llama-2-7b~\cite{touvron2023llama} & ONNX RT~\cite{onnxruntime} & 7B & 1.9 s\\ 
Stable Diffusion~\cite{rombach2022high} & TorchInductor~\cite{torchinductor2024} & 983M & 2.5 s\\ 
GPT-Neo~\cite{gpt-neo} & TorchInductor~\cite{torchinductor2024} & 2.7B & 3.6 s\\ 
\Xhline{2\arrayrulewidth}
\end{tabularx}
\vspace{5pt}
\caption{Training and inference benchmarks}
\label{benchmark-table}
\end{table}

\begin{figure*}[t]
  \centering
  \includegraphics[width=1\linewidth]{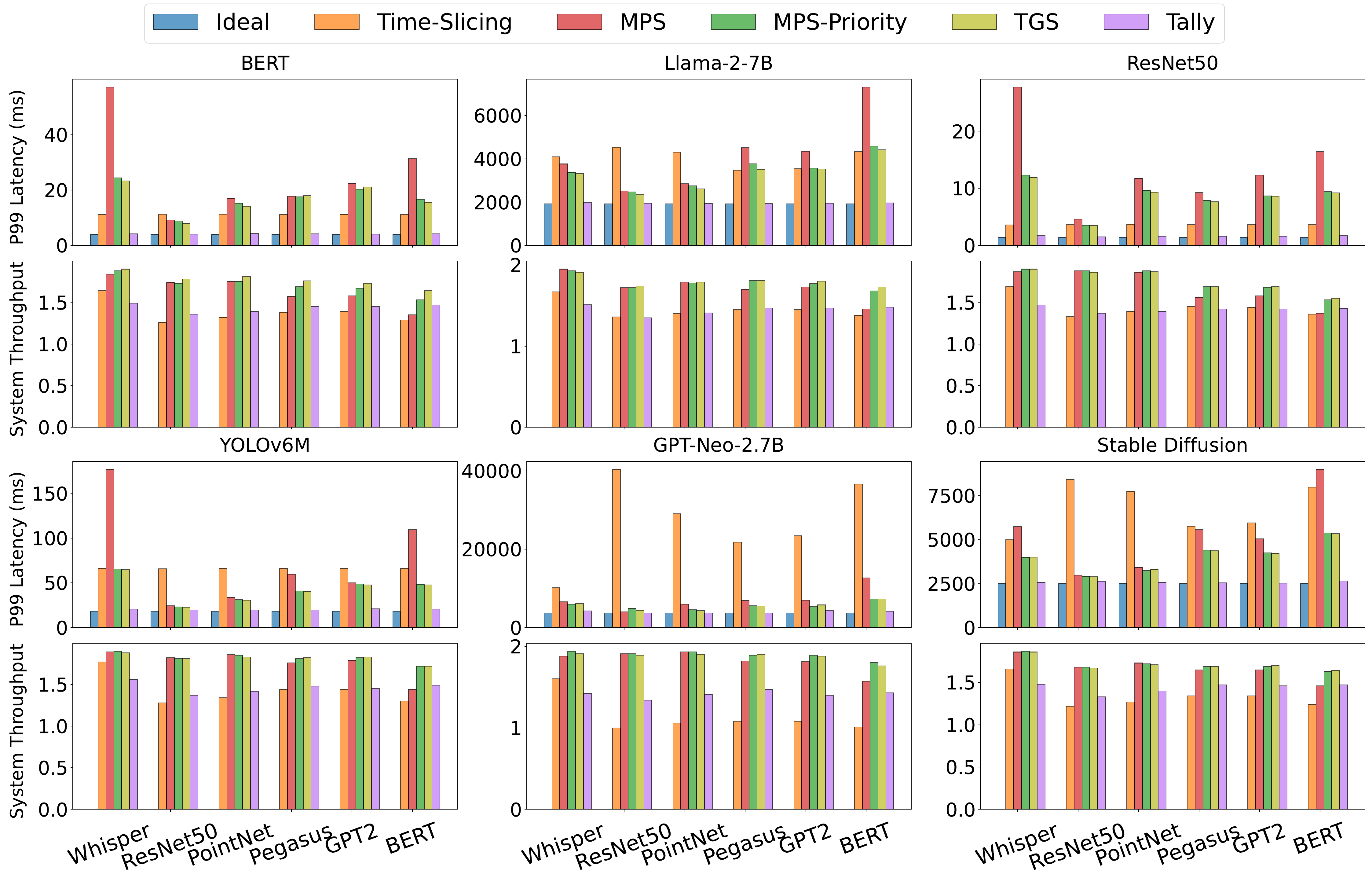}
  \caption{Analysis of 99th-percentile latency and system throughput across different high-priority inference and best-effort training workload combinations under various GPU sharing mechanisms.}
  \label{fig:end-to-end-eval}
\end{figure*}

\subsection{End-to-end Results}
\label{sec:end-to-end}
We first analyze Tally's end-to-end performance in efficiently sharing GPU resources among multiple workloads while guaranteeing performance isolation for high-priority inference tasks. To this end, we measure the $99^{th}$-percentile latencies of the inference tasks and the overall system throughput achieved by Tally when executed concurrently with a best-effort training workload. First, we compare the latency metrics against the state-of-the-art non-intrusive GPU sharing systems, namely Time-Slicing, MPS, MPS-Priority, and TGS, as well as against the ideal scenario of running the inference task in isolation. Figure~\ref{fig:end-to-end-eval} presents the latency and throughput measurements for all inference-training combinations from the benchmark suite, with inference tasks simulated using the MAF2 trace with a fixed $50\%$ load. The results demonstrate that Time-Slicing, MPS, MPS-Priority, and TGS exhibit significantly higher overhead compared to the ideal latency, with an average increase of $252.3\%$, $345.0\%$, $195.5\%$, and $188.9\%$, respectively. The suboptimal inference latencies exhibited by Time-Slicing and MPS can be attributed to their priority-agnostic design, which focuses solely on maximizing system utilization. MPS-Priority and TGS, on the other hand, exhibit improved latencies by prioritizing kernel execution of high-priority tasks over best-effort tasks. However, their coarse-grained kernel-level scheduling strategies fail to effectively mitigate co-execution interference. In stark contrast, Tally consistently maintains tail latencies nearly identical to the ideal latencies of the high-priority workloads, with an average percentage increase of mere $7.2\%$. This stronger performance isolation capability can be attributed to the block-level scheduling strategy that Tally employs.

In addition to latency, we also compare the overall system throughput of Tally against the baseline GPU sharing systems. Tally demonstrates competitive performance relative to the baselines, attaining an average of $105.2\%$, $83.6\%$, $80.6\%$, and $80.3\%$ of the system throughput achieved by the four baseline systems, respectively. This shows the ability of Tally's priority-aware scheduler to opportunistically schedule best-effort kernels while preserving latency requirements of high-priority tasks in shared execution environments.

\begin{figure*}[t]
    \centering
    \begin{subfigure}[b]{0.53\textwidth}
        \centering
        \includegraphics[width=\textwidth]{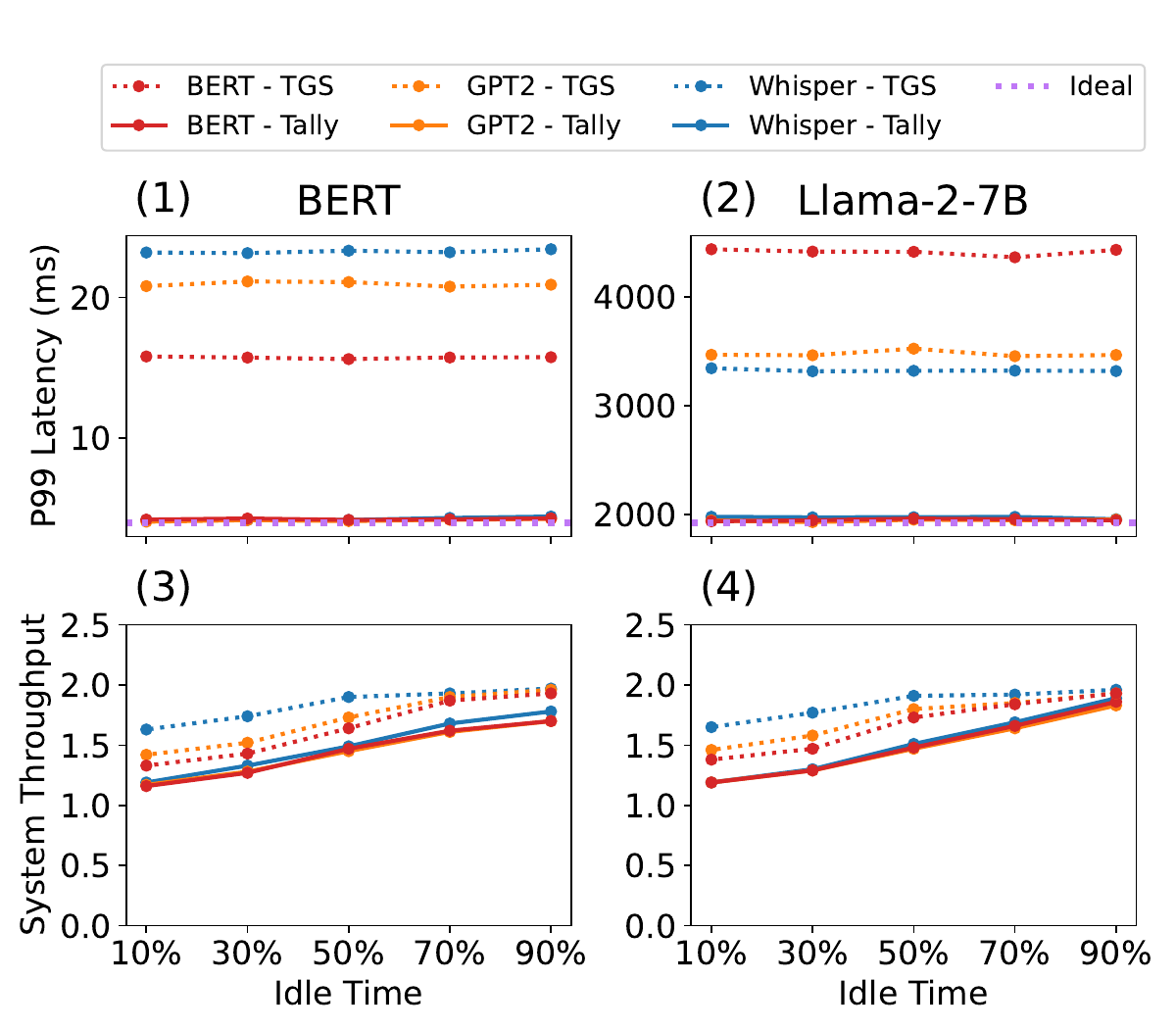}
    \end{subfigure}
    \hfill
    \begin{subfigure}[b]{0.44\textwidth}
        \centering
        \includegraphics[width=\textwidth]{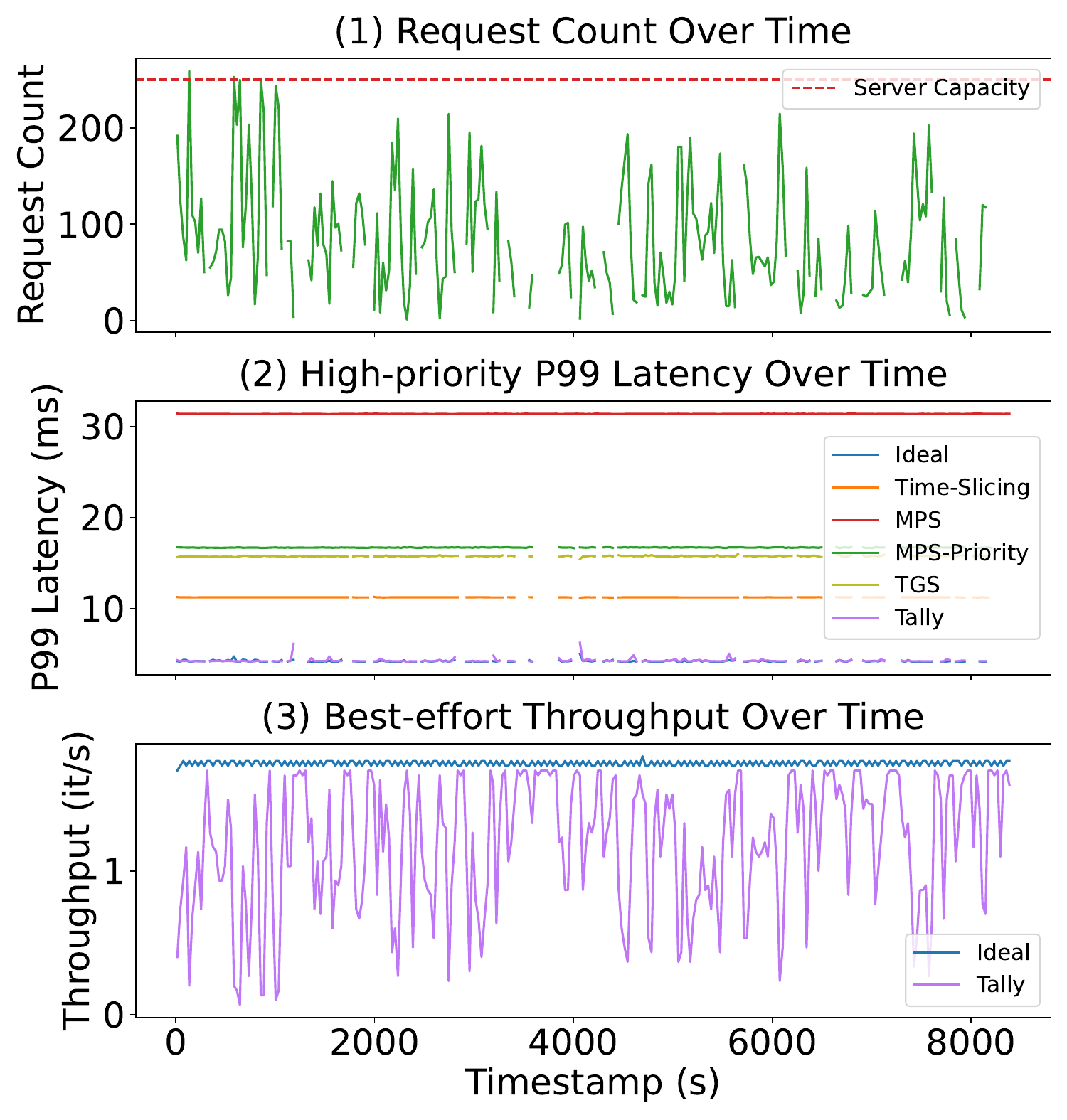}
    \end{subfigure}
    \caption{(a) Latency and throughput for high-priority BERT and Llama2-7B inference co-located with BERT, GPT-2, and Whisper training under Tally and TGS across different traffic loads. (b) Time-series visualization of user traffic, tail latency, and throughput over time for BERT inference co-located with BERT training across different GPU sharing systems. }
    \label{fig:real-time-eval-varying-load}
\end{figure*}

Furthermore, we observe that MPS, MPS-Priority, and TGS exhibit substantial variation in performance degradation for high-priority inference tasks across different workload combinations. For instance, TGS experiences performance degradation ranging from $15.6\%$ to $751.7\%$. This variation in latency reveals the sensitivity of these systems to workload characteristics, such as the differences in kernel duration across various tasks. As these systems schedule at the kernel level, their performance deteriorates more severely for workloads with longer kernels. In contrast, Tally's block-level scheduling demonstrates robust performance against variations in kernel duration, effectively limiting the performance degradation of inference tasks to under $10\%$ across $70\%$ of the workload combinations (worst-case $23\%$ on ResNet50). 

\subsection{Traffic Load Sensitivity Analysis}
We next evaluate the sensitivity of Tally's performance isolation capabilities to varying traffic loads. For this experiment, we use BERT and Llama-2-7B as the inference tasks representative of low-latency ($3.93$ ms) and long-latency ($1.9$ s) workloads. Each of them is paired with three distinct training workloads: BERT, GPT2, and Whisper, chosen for their observed tendency to cause significant interference. Figure~\ref{fig:real-time-eval-varying-load}a (1) and (2) present a comparative analysis of the $99^{th}$-percentile latency for the inference tasks under Tally and TGS relative to the original latency as a function of GPU idle time (calculated as $100\%$ - \emph{load}) ranging from $10\%$ to $90\%$. The results demonstrate that both inference tasks under Tally exhibit latencies indistinguishable from their original performance consistently across varying loads. In contrast, TGS incurs latency slowdowns of up to $5.8\times$ and $2.3\times$ for BERT and Llama-2-7B, respectively. These findings demonstrate the robustness of Tally's performance isolation capabilities for high-priority inference tasks, regardless of traffic load or the characteristics of the paired training workload.

We further analyze the relationship between Tally's achievable throughput and traffic load. Figure~\ref{fig:real-time-eval-varying-load}a (3) and (4) illustrate the system throughput under Tally and TGS as a function of idle time. Both systems exhibit an upward trend in system throughput as the percentage of idle time increases, indicating their ability to leverage available resources for improved performance. During periods of high traffic load, there is a noticeable difference between the two systems. For instance, at $10\%$ idle time, the system throughput under Tally is $1.19$, whereas TGS achieves $1.65$. This difference is attributed to Tally's strict priority enforcement policy, which executes best-effort tasks only when the high-priority task is inactive, sacrificing potential throughput gains for the preservation of the high-priority task's latency. In contrast, TGS achieves higher throughput but at the cost of significant slowdowns in tail latency, as demonstrated earlier. Nevertheless, as idle time increases, the throughput difference between Tally and TGS diminishes. This convergence indicates that for scenarios where heavy underutilization is observed, Tally not only offers reliable performance isolation guarantees but also achieves competitive throughput compared to other GPU sharing systems.

To illustrate the adaptive nature of Tally to varying traffic load, in Figure~\ref{fig:real-time-eval-varying-load}b, we plot the real-time traffic trace to a BERT inference task (a condensed version of the original MAF2 trace), and measure the real-time tail latency and throughput when co-located with another BERT training task. Figure~\ref{fig:real-time-eval-varying-load}b illustrates (1) the real-time traffic to a BERT inference task, (2) $99^{th}$ percentile latency of the inference server achieved by different GPU sharing systems and (3) the throughput of the co-located best-effort BERT training achieved by Tally compared to the original throughput of the training task. As shown in Figure~\ref{fig:real-time-eval-varying-load}b (2), the tail latency under Tally, represented by the purple line, closely aligns with the ideal latency depicted by the blue line, with negligible deviation throughout the simulation period. Whereas alternative approaches such as Time-Slicing, MPS, MPS-Priority, and TGS result in substantial slowdowns in latency as depicted. Additionally, Figure~\ref{fig:real-time-eval-varying-load}b (3) shows Tally's ability to opportunistically adjust the throughput of the best-effort task in correspondence with the fluctuating traffic load to the high-priority inference task, as discussed in Section~\ref{sec:tally-system}. As a result, Tally preserves over $68\%$ of the training task's original throughput simultaneously running the high-priority inference workload on the same GPU without affecting its latency requirements.

\begin{figure*}[t]
  \centering
  \includegraphics[width=1\linewidth]{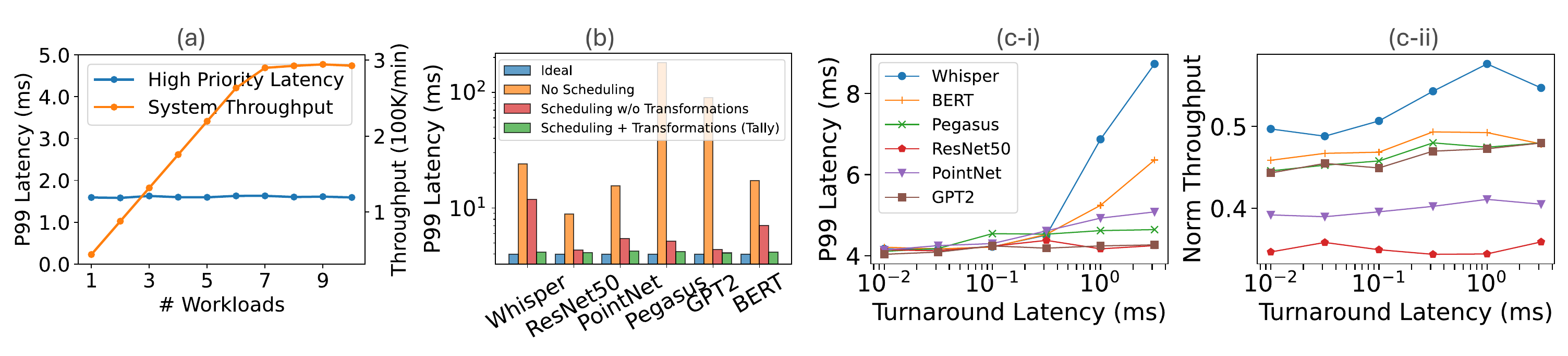}
  \caption{(a) Scalability performance of Tally with respect to number of concurrent workloads. (b) Performance decomposition of Tally for BERT inference. (c) Latency and throughput under different turnaround latency threshold settings.}
  \label{fig:extensive-exp}
\end{figure*}

\subsection{Scalability with Number of Workloads}

To demonstrate Tally's ability to support multiple best-effort workloads while preserving the performance of the co-located high-priority job, we extend our experiments beyond the two-job scenario. We designed an experiment with a set of identical ResNet50 inference workloads, each operating at a low traffic load of $10\%$ and one of these workloads designated as the high-priority online inference task while the rest as best-effort offline inference tasks. We measure $99^{th}$-percentile latency of the high-priority task and the aggregated system throughput, measured in number of requests processed per minute, as we scale up the number of best-effort tasks. The results in Figure~\ref{fig:extensive-exp}(a), reveal that the latency of the high-priority task remains consistent throughout the spectrum of scaling from $1$ to $10$ workloads. At the same time, the system throughput exhibits a steady increase as the number of workloads grows, until the GPU saturates at $8$ concurrent best-effort workloads. This shows that Tally can be particularly useful in scenarios where numerous jobs exhibit extremely low utilization, as Tally can efficiently pack these workloads together to maximize system throughput without compromising the performance of high-priority tasks.

\subsection{Performance Decomposition}
\label{sec:ablation}

We analyze the importance of Tally's priority-aware scheduling algorithm and block-level transformations for performance isolation. Figure~\ref{fig:extensive-exp}(b) compares the $99^{th}$-percentile latency of a high-priority BERT model when co-located with training workloads from the benchmark suite. \emph{Ideal} denotes the original latency of the task executed in isolation. \emph{No-scheduling} refers to the indiscriminate dispatching of kernels from both workloads upon arrival. \emph{Scheduling w/o Transformations} employs Tally's priority-aware scheduling without kernel transformations, while \emph{Scheduling with Transformations} represents the full implementation of Tally, utilizing slicing and preemption mechanisms for block-level scheduling. The results reveal that No-scheduling incurs substantial latency overhead across all $6$ job combinations, with the most severe case exhibiting a $30\times$ slowdown when co-located with Whisper. This significant performance degradation is attributable to the absence of priority enforcement, resulting in considerable co-execution interference. The integration of Tally's priority-aware kernel-level scheduling strategy somewhat mitigates this overhead. For ResNet50 and GPT2, this approach yields latencies close to ideal performance, with $8.0\%$ and $9.8\%$ slowdown respectively. However, for workloads such as Whisper and BERT, the latency slowdown can still reach up to $10\times$. This difference in slowdown is attributed to the variation in kernel duration distribution. For instance, $99.3\%$ of kernels in ResNet50 complete in less than $0.1$ ms, whereas $5.6\%$ of kernels in Whisper exceed the duration of an entire BERT inference ($3.93$ ms). These prolonged kernels significantly delay the execution of high-priority kernels. Notably, with kernel transformation enabled, i.e., with full Tally, the latency overhead becomes negligible across all six training workloads, resulting in an average of $4.0\%$ slowdown ($6.2\%$ in the worst case). This shows the necessity of block-level scheduling in maintaining robust performance isolation guarantees.

\subsection{Turnaround Latency Analysis}
\label{sec:turnaround_latency}
We examine the impact of the turnaround latency threshold (delineated in Section~\ref{sec:tally-system}) by evaluating the tail latency of the BERT inference workload and the normalized throughput of the co-located training job on six models across six distinct threshold values, ranging from $0.01$ ms to $10$ ms. The results, depicted in Figure~\ref{fig:extensive-exp} (c-i) and (c-ii), show that higher turnaround latency thresholds generally correspond to increased tail latency for the inference workload with only slight increase in the throughput. Based on these empirical results, we find that a turnaround latency threshold of $0.0316$ ms provides the best balance between latency and throughput, which is the default setting we use throughout experiments.

\subsection{Overhead Analysis}

\noindent \textbf{Virtualization.} We evaluate Tally's virtualization overhead by measuring the throughput of each workload in the benchmark suite and comparing it to their baseline performance. Results indicate that virtualization incurs an average overhead of only $1\%$, proving minimal performance impact.

\noindent \textbf{Kernel transformation.} We evaluate the overhead of Tally's kernel transformation by profiling $10K$ kernels from best-effort workloads (including same kernel with varying work dimensions). Our analysis reveals an average overhead of $25\%$. Notably, this overhead is exclusive to best-effort kernels. The performance of high-priority jobs remain unaffected.

\noindent \textbf{Profiling.} Tally profiles each new kernel from best-effort tasks to determine its optimal launch configuration. Once collected, these measurements are reused throughout the execution. Thus, Tally's profiling stage for a job generally completes within minutes. This overhead is negligible given that training workloads often run for hours to days.
\section{DISCUSSION}
\noindent \textbf{Comparison with MIG.} NVIDIA MIG~\cite{nvidia-mig} enables hardware-level GPU partitioning into isolated instances, providing robust performance isolation for concurrent workloads. However, MIG's coarse-grained partitioning mechanism inherently limits the maximum resource allocatable to individual workloads, which can result in potential performance degradation of high-priority tasks. This limitation stems from MIG's requirement for configuring resource partitions prior to task execution; any alteration to these partitions necessitates the termination of all running processes, rendering it impractical in production environments where job interruption is typically prohibited. Even when reconfiguration is feasible, this process can incur significant overhead: resetting partitions takes several seconds via command-line instructions, and applications require additional time from seconds to minutes to resume execution. As a result, MIG struggles to promptly adapt resource allocations to meet the dynamic demands of DL workloads, potentially leading to compromised performance of high-priority workloads. Additionally, MIG's static allocation scheme can result in suboptimal resource utilization, as idle cycles within one partition cannot be dynamically reallocated to others. These limitations prevent MIG from simultaneously achieving high utilization while maintaining the performance of high-priority tasks. In contrast, Tally effectively addresses these challenges by dynamically allocating resources based on real-time resource demands, allowing high-priority tasks to fully utilize the GPU when active while reallocating idle cycles to best-effort tasks, thereby ensuring both high utilization and effective performance isolation.

\noindent \textbf{Generalization to CUDA extensions.} While the standard GPU programming model enforces the independence of thread blocks within a kernel (as explained in Section~\ref{sec:bac}), recent CUDA extensions, such as \emph{Thread Block Clusters}~\cite{thread-block-clusters} and \emph{Cooperative Groups}~\cite{cooperative-groups}, introduce support for inter-block dependencies under certain constraints. We illustrate how Tally can reliably detect and handle these cases. (1) Thread Block Clusters organizes thread blocks into clusters, enabling intra-cluster communication via hardware support. These kernels can be identified by special flags passed at kernel launch. In such cases, Tally applies slicing and preemption at the cluster level, rather than at the thread block level. This ensures execution correctness while enabling more fine-grained scheduling. (2) Cooperative Groups introduces new thread grouping mechanisms that extend the capabilities of the standard CUDA programming model. Among them is inter-block synchronization, which enables coordinated execution across thread blocks. Kernels using this feature are launched through the \textit{cudaLaunchCooperativeKernel} API, making them easily identifiable. Notably, inter-block synchronization requires that all thread blocks within a kernel must be simultaneously scheduled on the GPU. For these kernels, Tally refrains from applying block-level scheduling to ensure compliance with the concurrent execution requirements. In our evaluation, none of the workloads employ inter-block cooperative groups, suggesting that the absence of block-level scheduling for these kernels has a negligible impact on Tally's overall performance isolation guarantees.
\section{RELATED WORK}
Many GPU sharing systems have been proposed with the goal of maximizing throughput. NVIDIA's Time-Slicing~\cite{GILMAN2021102234} and MPS~\cite{nvidia-mps}, for example, offer mechanisms to share GPU resources in temporal and spatial manners, respectively. Moreover, several studies~\cite{Ebird2019, gpupool2023, smgpu2016} have explored co-execution of kernels with complementary resource demands (e.g., compute vs. memory) to exploit resource heterogeneity in GPUs. Systems like Kernelet~\cite{zhong2013kernelet} and Slate~\cite{slate2019} schedule such co-execution dynamically at runtime, while approaches like Kernel Fusion~\cite{fusion2010}, HFUSE~\cite{li2022automatic}, Tacker~\cite{tacker2022}, and ISPA~\cite{ispa2023} employ source-to-source transformations to fuse kernels during the compilation phase. Despite their efficacy in improving resource utilization, these systems inherently lack support for performance isolation, as they prioritize overall system efficiency over preservation of performance of individual workloads, making them unsuitable for production environments where tasks must adhere to strict SLAs.

To address the need for performance isolation, solutions like Effisha~\cite{effisha2017} and GPES~\cite{gpes2015} utilize block-level preemption techniques to mitigate interference during concurrent execution. However, these approaches require access to the application source code, whereas Tally operates non-intrusively. Other solutions, such as AntMan~\cite{xiao2020} and TGS~\cite{tgs2023}, leverage the iterative nature of DL training to adaptively limit the resources of low-priority tasks based on performance feedback from high-priority jobs. Meanwhile, REEF~\cite{reef2022} implements a reset-based scheduling mechanism that achieves microsecond-scale kernel preemption latency on idempotent kernels. Yet, these works rely on specific task characteristics and cannot generalize to the wide variety of DL workloads. In contrast, Tally supports broad compatibility using task-agnostic scheduling and kernel transformations. Furthermore, Orion~\cite{orion2024} enhances performance isolation via kernel-level scheduling with consideration on each kernel's compute and memory requirements. However, it requires offline profiling of workloads, which is challenging to accomplish in cluster settings. In comparison, Tally employs transparent profiling for priority-aware scheduling.
\section{CONCLUSION}
In this paper, we highlight the limitations of current GPU sharing systems and their inability to meet the diverse requirements of modern large-scale DL clusters. To overcome these challenges, we design Tally, a novel non-intrusive GPU sharing mechanism that offers robust performance isolation while maintaining compatibility with a wide range of DL workloads. Our evaluation shows that Tally can effectively ensure the performance standards of latency-critical tasks during shared execution, incurring a mere $7.2\%$ overhead on average while offering competitive system throughput.

\begin{acks}
We thank the anonymous reviewers and our shepherd, Tim Rogers, for their valuable feedback. This research was supported in part by the Innovation JELF Grant, NSERC Discovery Grant, AWS Machine Learning Research Award (MLRA), Facebook Faculty Research Award, Google Scholar Research Award, and VMware Early Career Faculty Grant.
\end{acks}

% LaTeX template for Artifact Evaluation V20240722
%
% Prepared by Grigori Fursin with contributions from Bruce Childers,
%   Michael Heroux, Michela Taufer and other colleagues.
%
% See examples of this Artifact Appendix in
%  * ASPLOS'24 "PyTorch 2: Faster Machine Learning Through Dynamic Python Bytecode Transformation and Graph Compilation": 
%      https://dl.acm.org/doi/10.1145/3620665.3640366
%  * SC'17 paper: https://dl.acm.org/citation.cfm?id=3126948
%  * CGO'17 paper: https://www.cl.cam.ac.uk/~sa614/papers/Software-Prefetching-CGO2017.pdf
%  * ACM ReQuEST-ASPLOS'18 paper: https://dl.acm.org/citation.cfm?doid=3229762.3229763
%
% (C)opyright 2014-2024 cTuning.org
%
% CC BY 4.0 license
%

% \documentclass{sigplanconf}

% \usepackage{hyperref}

% \begin{document}

% \special{papersize=8.5in,11in}

%%%%%%%%%%%%%%%%%%%%%%%%%%%%%%%%%%%%%%%%%%%%%%%%%%%%
% When adding this appendix to your paper, 
% please remove above part
%%%%%%%%%%%%%%%%%%%%%%%%%%%%%%%%%%%%%%%%%%%%%%%%%%%%

\appendix
\section{Artifact Appendix}

%%%%%%%%%%%%%%%%%%%%%%%%%%%%%%%%%%%%%%%%%%%%%%%%%%%%%%%%%%%%%%%%%%%%%
\subsection{Abstract}

We provide the source code of Tally along with scripts to reproduce experimental results presented in the evaluation section. This appendix includes instructions for generating plots similar to Figure~\ref{fig:end-to-end-eval} and Figure~\ref{fig:extensive-exp}(b). Dockerfiles are provided to set up the required runtime environment. To expedite artifact evaluation, a pre-built image is available, containing the fully configured environment, pre-compiled deep learning models, and datasets as specified in Table \ref{benchmark-table}. The experiments require an x86-64 Linux host with at least 85 GB RAM, 200 GB of free disk space, and an NVIDIA A100 GPU (40 GB). For consistent performance, we recommend using a Google Cloud a2-highgpu-1g instance.

\subsection{Artifact check-list (meta-information)}

{\small
\begin{itemize}
  \item {\bf Run-time environment:} Docker.
  \item {\bf Compilation:} Pre-compiled within a Docker image.
  \item {\bf Data set:} MAF2 trace and datasets as listed in Table~\ref{benchmark-table}.
  \item {\bf Hardware:} A 12-core CPU with 85 GB RAM and an NVIDIA A100 GPU (40 GB).
  \item {\bf Metrics:} $99^{th}$ percentile latency and throughput.
  \item {\bf Experiments: } End-to-end evaluation (Section~\ref{sec:end-to-end}) and performance decomposition (Section~\ref{sec:ablation}). For faster evaluation, the default setup generates results only for co-executing BERT and Llama inference tasks alongside all training workloads. Instructions for running the full set of inference workloads can be found in Section~\ref{sec:exp-customize}.
  \item {\bf Output: } Plots similar to Figure~\ref{fig:end-to-end-eval} and Figure~\ref{fig:extensive-exp}(b).
  \item {\bf How much disk space required (approximately)?:} Approximately 200 GB.
  \item {\bf How much time is needed to prepare workflow (approximately)?:} About 10 minutes to download the Docker image.
  \item {\bf How much time is needed to complete experiments (approximately)?: } Approximately 6 hours for BERT and Llama tasks; 18 hours for the full end-to-end evaluation.
  \item {\bf Publicly available?: } Yes.
  \item {\bf Code licenses (if publicly available)?: } MIT License.
\end{itemize}
}

%%%%%%%%%%%%%%%%%%%%%%%%%%%%%%%%%%%%%%%%%%%%%%%%%%%%%%%%%%%%%%%%%%%%%
\subsection{Description}

\subsubsection{How to access} The artifact is available for download on GitHub: \href{https://github.com/tally-project/tally-bench}{https://github.com/tally-project/tally-bench}.

\subsubsection{Hardware dependencies} Requires an x86-64 Linux host with at least 85 GB of RAM, 200 GB of free disk space, and an NVIDIA A100 GPU (40 GB). We recommend a Google Cloud a2-highgpu-1g instance for consistent performance.

\subsubsection{Software dependencies} Docker installation is required as all experiments are configured to run within the provided Docker image.

\subsubsection{Data sets} The data sets used in the experiments are listed in Table~\ref{benchmark-table} and are pre-downloaded as part of the Docker image.

\subsubsection{Models} The models used in these experiments are listed in Table~\ref{benchmark-table}.

%%%%%%%%%%%%%%%%%%%%%%%%%%%%%%%%%%%%%%%%%%%%%%%%%%%%%%%%%%%%%%%%%%%%%
\subsection{Installation} 

\begin{itemize}
    \item[1.] Clone the Github repository.
{\footnotesize
\begin{verbatim}
$ git clone
  https://github.com/tally-project/tally-bench.git
$ cd tally-bench
\end{verbatim}
}
    \item[2.] Pull the Docker image.
{\footnotesize
\begin{verbatim}
$ docker pull wzhao18/tally:bench
\end{verbatim}
}
\end{itemize}

%%%%%%%%%%%%%%%%%%%%%%%%%%%%%%%%%%%%%%%%%%%%%%%%%%%%%%%%%%%%%%%%%%%%%
\subsection{Experiment workflow}

\begin{itemize}
    \item[1.] Create the results directory and launch the Docker container.
{\footnotesize
\begin{verbatim}
$ mkdir tally_results
$ docker run -it --shm-size=64g --gpus 0 -v
  ${PWD}/tally_results:/home/tally-bench/tally_results
  wzhao18/tally:bench /bin/bash
\end{verbatim}
}
    \item[2.] Run the following commands to run the experiments.
{\footnotesize
\begin{verbatim}
(On host machine)
$ sudo nvidia-smi -i 0 -c DEFAULT

(In docker container)
$ ./scripts/run_bench.sh > tally_results/b1.log 2>&1

(On host machine)
$ sudo nvidia-smi -i 0 -c EXCLUSIVE_PROCESS

(In docker container)
$ ./scripts/run_bench.sh > tally_results/b2.log 2>&1
\end{verbatim}
}
\end{itemize}

%%%%%%%%%%%%%%%%%%%%%%%%%%%%%%%%%%%%%%%%%%%%%%%%%%%%%%%%%%%%%%%%%%%%%
\subsection{Evaluation and expected results}
\begin{itemize}
    \item[1.] Parse results into CSV files.
{\footnotesize
\begin{verbatim}
(On host or in container)
$ python3 ./scripts/parse_results.py
\end{verbatim}
}
    \item[2.] Generate plots.
{\footnotesize
\begin{verbatim}
(On host or in container)
$ python3 ./scripts/plot_results_micro.py
\end{verbatim}
}
    \item[3.] Plots similar to Figure~\ref{fig:end-to-end-eval} and Figure~\ref{fig:extensive-exp}(b) can be found in the \texttt{tally\_results} directory.

\end{itemize}

%%%%%%%%%%%%%%%%%%%%%%%%%%%%%%%%%%%%%%%%%%%%%%%%%%%%%%%%%%%%%%%%%%%%%
\subsection{Experiment customization}
\label{sec:exp-customize}
To run the full benchmark, pass the \verb|--run-full-benchmark| flag to \verb|run_bench.sh|. Use the \texttt{scripts/plot\_results.py} script to generate the plots. Additional scripts for running other experiments and producing figures included in the paper are available in the \texttt{scripts} directory.

%%%%%%%%%%%%%%%%%%%%%%%%%%%%%%%%%%%%%%%%%%%%%%%%%%%%
% When adding this appendix to your paper, 
% please remove below part
%%%%%%%%%%%%%%%%%%%%%%%%%%%%%%%%%%%%%%%%%%%%%%%%%%%%

% \end{document}

\bibliographystyle{plain}
\balance
\bibliography{references}

\end{document}